\documentclass[fleqn,10pt]{wlscirep}
\usepackage{amsmath}
\usepackage{bm}
\title{Hartman effect for spin waves in exchange regime}

\author[1,2,*]{Jarosław W. Kłos}
\author[3,4]{Yuliya S. Dadoenkova}
\author[1]{Justyna Rychły}
\author[3,4]{Nataliya~N.~Dadoenkova}
\author[4,5]{Igor L. Lyubchanskii}
\author[1,6]{Józef Barnaś}
\affil[1]{Faculty of Physics, Adam Mickiewicz University in Pozna\'{n}, 61-614 Pozna\'{n}, Poland}
\affil[2]{Institute of Physics, Greifswald University,
17489 Greifswald, Germany}
\affil[3]{Ulyanovsk State University, 432017 Ulyanovsk, Russia}
\affil[4]{Donetsk Physical and Technical Institute of the National Academy of Sciences of Ukraine, Ukraine}
\affil[5]{Faculty of Physics, V. N. Karazin Kharkiv National University, 61022 Kharkiv, Ukraine}
\affil[6]{Institute of Molecular Physics, Polish Academy of Sciences, 60-179 Poznań, Poland}

\affil[*]{klos@amu.edu.pl}


\begin{abstract}
Hartman effect for spin waves tunnelling through a barrier in a thin magnetic film is considered theoretically. The barrier is assumed to be created by a locally increased magnetic anisotropy field. The considerations are focused on a nanoscale system operating in the exchange-dominated regime. We derive the formula for group delay $\tau_{gr}$ of a spin wave packet and show that $\tau_{gr}$ saturates with increasing barrier width, which is a signature of the Hartman effect predicted earlier for photonic and electronic systems. In our calculations we consider the general boundary conditions which take into account different strength of exchange coupling between the barrier and its surrounding. As a system suitable for experimental observation of the  Hartman effect we propose a CoFeB layer with perpendicular magnetic anisotropy induced by a MgO overlayer.
\end{abstract}
\begin{document}

\flushbottom
\maketitle
%
%
\thispagestyle{empty}


\section*{Introduction}

The problem of quantum tunneling of a particle  through a potential barrier higher than the particle energy is one of the fundamental problems in quantum mechanics~\cite{Landau,Messiah,Roy,Olkhovsky04}.  More than half a century ago Hartman considered analytically tunneling of a Gaussian wave packet through a rectangular potential barrier of thickness $L$ in metal/insulator/metal junctions~\cite{Hartman}. He derived a formula for the group delay $\tau_{gr}$, i.e. the time in which the incident packet travels from the first border of the barrier at $x=0$ to the second one at $x=L$. He concluded that the group delay $\tau_{gr}$ saturates with increasing barrier thickness, which means that for thick barriers the group delay is shorter than the time required by the packet to traverse the distance $L$ in the corresponding uniform (no barrier) material. This phenomenon is known as the Hartman effect (HE).

However, tunneling is a wave phenomenon  which refers to various kinds of waves, including also  wave functions of quantum particles. In general, the tunneling takes place when the wave passing through the barrier has the form of exponentially evanescent function, and then continues the propagation as an ordinary wave (in oscillatory form) with reduced amplitude and shifted phase. The Hartman's paper initiated a huge activity in different fields: including (i) tunneling of electromagnetic Gaussian wave packets in various structures like photonic crystals (see, for example, review articles~\cite{Olkhovsky,Winful03,Winful03b,Schwartsburg07,Niemtz97,Olkhovsky04,Niemtz03,Olkhovsky14} and research papers~\cite{Winful03,Wang04,Sahrai15,Dadoenkova16,Jamil17}), (ii) tunneling of acoustical  and optical phonons\cite{Huynh06,Villegas17}, and (iii) tunneling of electrons in graphene~\cite{Wu09,Sepkhnov09,Park14,Chen14,Ban15}. For all types of the above mentioned waves, the saturation of group delay with increasing barrier width is observed in tunneling processes. This  feature leads to a counter-intuitive conclusion on {\it an unlimited increase of  propagation speed} of tunneling wave packets. This paradox was the subject of intensive scientific debate\cite{Winful06} and was explained using the arguments referring to a reshaping of the wave packets\cite{Buttiker03} or to the saturation of energy deposition within the barrier\cite{Winful02}.

The HE was not studied yet in the case of spin waves (SWs), although the effects related to tunneling, trapping and mastering of propagation time or velocity for SWs in non-uniform magnetic structures have been already investigated theoretically and experimentally. Most of experimental studies were performed for dipolar SWs in structures based on yttrium-iron garnet (YIG) or on permalloy. Tunneling of SW was investigated experimentally in YIG stripes with a single\cite{Demokritov04} (or double\cite{Hansen07}) barrier formed by the Oersted field or in YIG film with a mechanical gap \cite{Schneider10}. There are also reports which show that the group delay or group velocity for SWs  in periodic structures can be significantly changed in comparison  to those in homogeneous systems\cite{Bankowski15,Chumak12,Neusser11,Tacchi15}.

In this paper we consider HE for SWs tunneling through a  barrier in a thin magnetic film with perpendicular magnetic anisotropy (PMA). We restrict our considerations to an exchange-dominated region of the spin wave spectrum. We demonstrate theoretically saturation of the group delay for exchange SWs with increasing width of the barrier, which is an evident signature of HE.
The \textit{barrier} 
can be created by a local increase of the internal field, which can be caused by a change (increase) of the magnetocrystalline anisotropy  within the barrier. Such a barrier can be formed, for instance,  using a material with anisotropy higher than that in the remaining (left and right) parts of the junction (referred to in the following as \textit{matrix}).
However, to reduce spin wave scattering at the barrier/matrix interfaces, one can take a uniformly magnetized thin film of a material with low damping, and then, with etching techniques, fabricate a narrow stripe of reduced thickness. By covering the film with an insulating material, one can induce an interface anisotropy, which in a narrow stripe can be different (enhanced) from that in the other parts of the structure. Indeed, for a layer (up to a few nanometers thick) the main contribution to the effective magnetic anisotropy originates from surfaces and/or interfaces, which grows with decreasing layer thickness. More details on the system proposed for experimental investigations of HE are given in next section.

For dipolar spin waves we can form the barrier in a few other ways. If we consider the magnetic stripe of high magnetization saturation surrounded by two magnetic half-planes made of material of lower magnetization saturation then in the range between FMR frequencies of this two materials the spin wave tunneling can be observed -- the Demon-Eshbach modes will be evanescent in the barrier (stripe) and oscillating in the matrix (surrounding material). We suspect that for such system the Hartman effect should be also observed. The other the magnonic systems in which the Hartman effect is expected to be found are the structures with air gaps where the exponentially decaying magnetostacic potential (and associated dynamic demagnetizing field) can couple the spin wave dynamics across the air gap. The magnonic systems operating in dipolar regime are, in most cases, much easier for fabrication than the extremely small systems working in exchange regime. However, the theoretical description dipolar system is more challenging. Therefore we decided to start, in this paper, from the theoretical and numerical studies of Hartman effect for exchange waves. 

The paper is organized as follows. The system under consideration is described in section "The Model". In the next section, we present a theoretical description. In this section we discuss propagation of exchange SWs in the case of  spatially dependent anisotropy field. We also outline the theoretical basis of the HE for SWs. Numerical results and their discussion are presented in the subsequent section. The last section contains the summary and final conclusions. The manuscript is supplemented by the discussion of boundary conditions for SWs in exchange regime and technical details concerning derivation of the transmissivity function.

\section*{The Model}

\begin{figure}[ht]
\centering
\includegraphics[width=0.8\columnwidth]{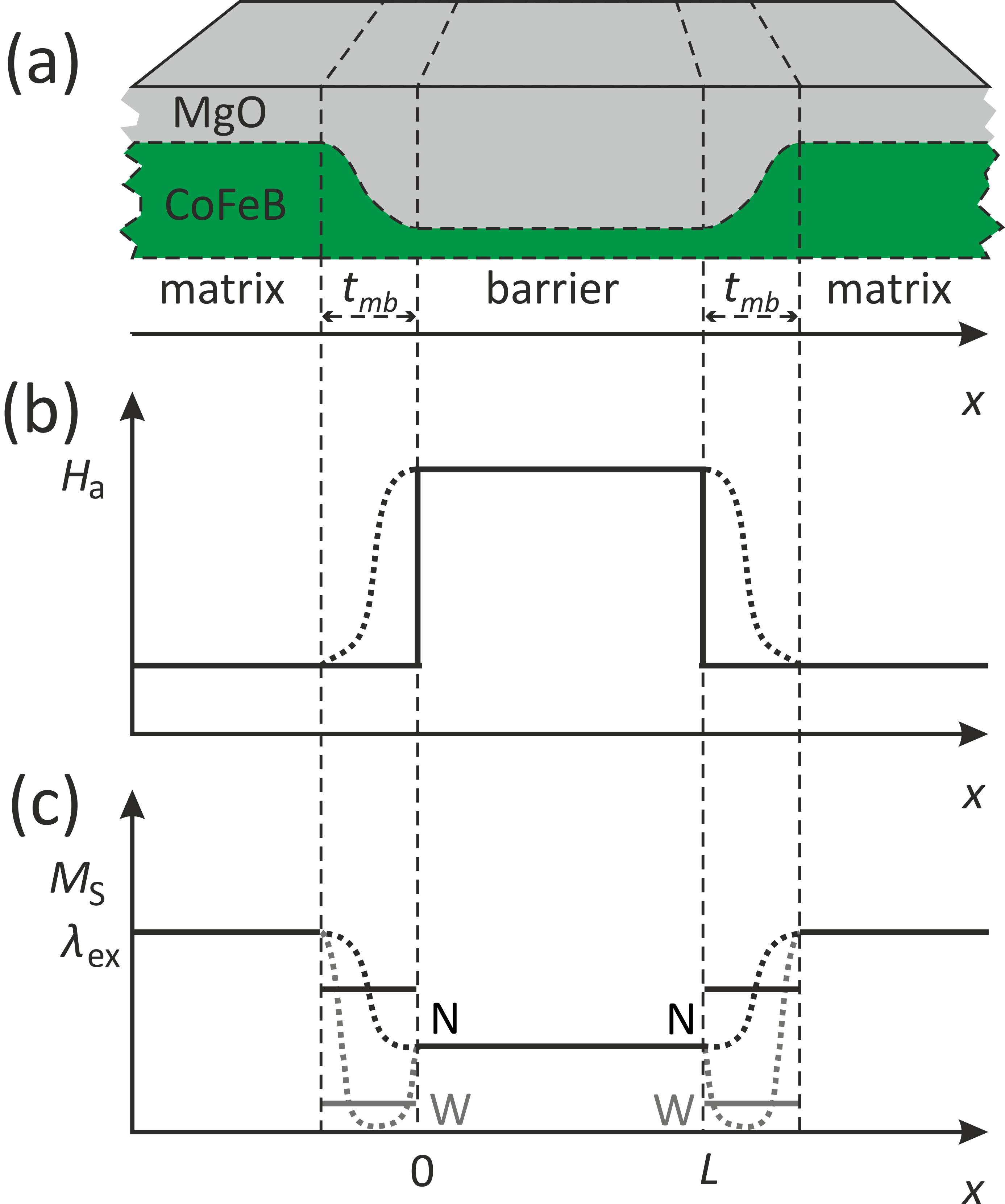}
\caption{Schematic of the system under consideration. (a) The exemplary structure has the form of a ferromagnetic layer made  of a low-damping material (CoFeB) with an out-of-plane magnetic anisotropy induced by the interface with the oxide layer (MgO) deposited on top of the ferromagnetic layer. The groove in the ferromagnetic layer makes  the effective anisotropy field $H_{\rm a}$ higher, forming the barrier (b) due to a larger contribution of the CoFeB/MgO interface anisotropy to $H_{\rm a}$. The change in thickness of the magnetic layer can also modify other material parameters (c): saturation magnetisation $M_{\rm S}$ and exchange length $\lambda_{\rm ex}$. In the calculations we used a simplified model with abruptly changing material parameters (solid lines). We also assumed that the material parameters at the interfaces between the barrier and matrix ($M_{\rm S,mb}$, $\lambda_{\rm ex,mb}$) can be different from the bulk parameters, and can correspond to the weak (W) or natural (N)  exchange coupling between the barrier and the matrix.
}
\label{fig:Fig1}
\end{figure}

The system under consideration, presented schematically in Fig.\ref{fig:Fig1}a, is planar. Both external magnetic field $\bm{H}_0$ and effective anisotropy field $\bm{H}_{\rm a}(x)$ are oriented out-of-plane. We assume a one-dimensional magnetic barrier in the form of a stripe region, in which (for $0<x<L$) the effective anisotropy field $H_{\rm a}(x)$ is increased. The barrier is rectangular, i.e. $H_{\rm a}$ changes abruptly at $x=0$ and $x=L$, see Fig.\ref{fig:Fig1}b. We also assume that the magnetization $M_{\rm S}$ and exchange length  $\lambda_{\rm ex}$ in the barrier region are changed (reduced) in reference to the matrix regions ($x<0$ and $x>L$), see Fig.\ref{fig:Fig1}c.

Thickness of the magnetic layer is much smaller than the considered wavelength of SWs and also smaller than the width of the barrier $L$. This allows to neglect the spatial changes of spin wave amplitude across the magnetic layer (regardless of the partial pinning which exists due to the interfacial anisotropy $K_{i}$ between the magnetic layer and nonmagnetic overlayer). To simplify our analysis, we investigateSWs propagating perpendicularly to the barrier only, which effectively reduces the problem to the one-dimensional one.

To take into account the in-plane inhomogeneity of the magnetic material, which in real systems can be observed at the interfaces between the matrix and barrier, we introduce the exchange coupling at these interfaces as an additional parameter of our model. Strength of this exchange coupling
is important for determining the interfacial boundary conditions for SWs. These boundary conditions significantly influence the phase factor of the transmissivity $T(\omega)$~\cite{Dadoenkova12,Gotte12,Gruszecki14,Gruszecki15}, which, in turn, is crucial for determination of the group delay $\tau_{\rm gr}(\omega)$ of SWs tunneling (propagating) through (over) the barrier~\cite{Winful06}. Therefore, one can expect that the HE (i.e. saturation of $\tau_{\rm gr}$ with increasing width of the barrier $L$) is sensitive to a particular formulation of the boundary conditions for SWs.

For a prospective experimental realization of the considered system we need a material characterized by a high out-of-plane anisotropy field, which additionally ensures low SW damping. A suitable system with PMA is a thin CoFeB layer covered with MgO. The effect of PMA induced at the CoFeB/MgO interface is well known and was already used in spintronics for fabrication of magnetic tunnel junctions of reduced dimensions\cite{Ikeda10}. The interfacial anisotropy $K_{i}$
depends critically on the crystalline structure and bonding at the CoFeB/MgO interfaces. It grows initially with increasing thickness of the MgO layer and then decreases for larger  thicknesses, when crystalline MgO starts to form\cite{Zhu14}.
For positive values of the energy density (effective anisotropy) $K=K_{i}/t_{\rm CoFeB}-\mu_0 M_{\rm S}^2/2$, the magnetic easy-axis is oriented out-of-plane and the system is magnetized perpendicularly in the absence of external field. Thus, by appropriate tuning of the MgO thickness and of the CoFeB thicknes ($t_{\rm CoFeB}$), one can increase the effective anisotropy inside the stripe region and form the barrier (see Fig.\ref{fig:Fig1}).

In the following sections we will consider propagation of exchange SWs in a nonuniform profile of effective (out-of-plane) anisotropy field $H_{\rm a}=2K_{i}/(\mu_{0}M_{\rm S}t_{\rm CoFeB})-M_{\rm S}$, additionally shifted by a spatially homogeneous external field $H_0$ (applied in the same direction). We will show that the field $H=H_{\rm a}(\bm{r})+H_{0}$ can be treated as a counterpart of electrostatic potential $V(\bm{r})$ for electronic waves. In the \textit{magnetic barrier}, where the spin wave frequency $\omega$ fulfills the condition $\gamma\mu_0\omega<H_{0}+H_{\rm a}(\bm{r})$, the spin wave profile has evanescent character, typical for tunnelling of electronic waves of energy $E<V(\bm{r})$. We will exploit this formal similarity of electronic waves and exchange SWs to discuss HE for magnonics.

\section*{Theoretical description}

In order to discuss HE for exchange dominated the SWs, we  start from  derivation of an  analytic formula for the transmissivity function $T(\omega)$ through a magnetic barrier embedded in a magnetic matrix, taking into account different types of boundary conditions at the barrier/matrix interfaces. Then, we derive the formula for the group delay $\tau_{gr}$ for spin wave packet tunneling (propagating) through (over) the magnetic barrier. Finally, we discuss the HE for exchange dominated SWs by analyzing the formula for group delay in the limit of wide barriers.

\subsection*{Exchange spin waves in spatially dependent anisotropy field}

In general, the magnetization dynamics in a magnonic system is described by the Landau-Lifshitz equation, which has the following form in the absence of damping:
\begin{equation}
\frac{\partial \bm{M}(\bm{r},t)}{\partial t}=-\gamma\mu_0 \bm{M}(\bm{r},t)\times\bm{H}_{\rm eff}(\bm{r},t),\label{eq:LLE}
\end{equation}
where $\bm{M}$ and $\bm{H}_{\rm eff}$ stand for the magnetization and the effective magnetic field, respectively, and $\gamma$ is the gyro-magnetic ratio. We consider a system in which the SWs of short wavelengths propagate in a magnetic layer with spatially varying (along the $x$-direction) material parameters: saturation magnetization  $M_{\rm S}(x)$, magnetic anisotropy $H_{\rm {a}}(x)$,  and exchange length $\lambda_{\rm{ex}}(x)$.  We assume that the effective field, $\bm{H}_{\rm{eff}}=\bm{H}_{\rm{0}}+\bm{H}_{\rm{a}}+\bm{H}_{\rm{ex}}$, includes the contributions from a uniform static external magnetic field $\bm{H}_0=[0,0,H_0]$, static and spatially dependent effective anisotropy field $\bm{H}_a(x)=[0,0,H_a(x)]$, and the dynamical term due to the exchange interaction between magnetic moments, $\bm{H}_{\rm{ex}}(x,t)$. The latter term can be written as\cite{Krawczyk12}:
\begin{equation}
\bm{H}_{\rm{ex}}(x,t)=\nabla \lambda_{\rm{ex}}^2(x)\nabla\bm{M}(x,t),
\end{equation}
where the magnetization $\bm{M}(x,t)$ precesses around the effective field $\bm{H}_{\rm eff}$,   $\bm{M}(x,t)\approx[m_{x}(x)e^{i\omega t},m_{y}(x)e^{i\omega t},M_{\rm{S}}]$.

When  considering propagation of SWs in the $x$-direction (normal to the barrier), the linearized Landau-Lifshits equation can be written in the form:
\begin{eqnarray}
i\frac{\omega}{\gamma\mu_0}m_x(x)&=&M_{\rm{S}}(x)\frac{\partial}{\partial x}\lambda_{\rm{ex}}^2(x)\frac{\partial}{\partial x}m_y(x)-m_y(x)\left(\frac{\partial}{\partial x}\lambda_{\rm{ex}}^2(x)\frac{\partial}{\partial x} M_{\rm{S}}(x)+H_0+H_{\rm a}(x)\right),\nonumber\\
 i\frac{\omega}{\gamma\mu_0}m_y(x)&=&-M_{\rm{S}}(x)\frac{\partial}{\partial x}\lambda_{\rm{ex}}^2(x)\frac{\partial}{\partial x}m_x(x)+m_x(x)\left(\frac{\partial}{\partial x}\lambda_{\rm{ex}}^2(x)\frac{\partial}{\partial x} M_{\rm{S}}(x)+H_0+H_{\rm a}(x)\right)
  .\label{eq:LLlin}
\end{eqnarray}
Substituting: $m_{+}(x)=m_{x}(x)+im_{y}(x)$ and $m_{-}(x)=m_{x}(x)-i\,m_{y}(x)$ we can write
Eqs.~(\ref{eq:LLlin}) in the following, more compact form:
\begin{equation}
-\frac{d}{dx}\lambda_{\rm{ex}}^2(x)\frac{d}{dx}m_{\pm }(x)+v(x)m_{\pm}(x)=M^{-1}_{\rm{S}}(x)\left(\pm\frac{\omega}{\gamma\mu_0}\right)m_{\pm}(x),\label{eq:schrod}
\end{equation}
where:
\begin{equation}
v(x)=M^{-1}_{\rm{S}}(x)\left(H_{0}+H_{a}(x)+\frac{d}{dx}\lambda_{\rm{ex}}^2(x)\frac{d}{dx}M_{\rm{S}}(x)\right).\label{eq:pot}
\end{equation}
Equation~(\ref{eq:schrod}) has the mathematical form of Sturm-Liouville equation, and therefore it possesses the properties of other differential equations of similar kind (e.g. of the Schr{\"o}dinger equation). One  can identify $\lambda_{ex}^2(x)$ and $v(x)$ as counterparts of the inverse effective mass and the effective potential, respectively. The last term in Eq.~(\ref{eq:pot}) contributes to the {\it effective potential} $v(x)$ only at the interfaces, at which the material parameters ($\lambda_{\rm{ex}}^2$, $M_{S}$) change.
The formal similarity of Eq.~(\ref{eq:schrod}) to Schr{\"o}dinger equation allows one to expect the HE for exchange SWs tunneling through a barrier, as well.

To find the solution of Eq.~(\ref{eq:schrod}) in the whole system (see Fig.\ref{fig:Fig1}), one has to match the solutions in homogeneous materials of the barrier and of the surrounding medium (matrix). For the barrier and matrix one can write  Eq.~(\ref{eq:schrod}) for $m_{+}$as:
\begin{eqnarray}
 -\tilde{M}_{\rm{S,\alpha}}\lambda_{\rm{ex},\alpha}^{2}\frac{d^2}{dx^2}m_{+}(x)+\left(1+\tilde{H}_{\rm{a,\alpha}}\right)m_{+}(x)=\Omega\, m_{+}(x),\label{eq:schrod2}
\end{eqnarray}
where $\alpha=\{\rm m,b\}$ refers to the matrix $\rm(m)$ or barrier $\rm(b)$, respectively.
The exchange lengths $\lambda_{\rm{ex},\rm m}$ in the matrix and $\lambda_{\rm{ex},\rm b}$ in the barrier  are measured in the units of spatial coordinate $x$. In turn,
$\tilde{M}_{\rm{S,\alpha}}$ and $\tilde{H}_{\rm{a,\alpha}}$ denote the dimensionless saturation magnetization and effective anisotropy field, respectively, for the matrix  or barrier:
\begin{equation}
\tilde{M}_{\rm{S,\alpha}}=\frac{M_{\rm{S,\alpha}}}{H_{0}},\,\,\tilde{H}_{\rm{a,\alpha}}=\frac{H_{\rm{a,\alpha}}}{H_{0}}\textcolor{blue}{,}
\end{equation}
whereas $\Omega$ is the dimensionless frequency:
\begin{equation}
\Omega=\frac{\omega}{\gamma\mu_0 H_{0}}.
\end{equation}
The general solution of Eq.~(\ref{eq:schrod2}) takes the form:
\begin{equation}
m_{+}(x)=C_{1}e^{i k_{\alpha}x}+C_{2}e^{-i k_{\alpha}x}\label{eq:gensol},
\end{equation}
where $C_{1}$ and $C_{2}$ are certain integration constants, while  $k_{\alpha}$ is the wave number which can be written in the form:
\begin{equation}
k_{\alpha}(\Omega)=\lambda_{\rm{ex,\alpha}}^{-1}\tilde{M}_{\rm{S,\alpha}}^{-\frac{1}{2}}\sqrt{\Omega-\left(1+ \tilde{H}_{\rm{a,\alpha}} \right)}.\label{eq:wavenum}
\end{equation}

\subsection*{Transmissivity and group delay}

Let's consider now the incident SW  $e^{i\left(k_{m}x+\Omega t\right)}$ of  frequency $\Omega>1+\tilde{H}_{\rm a,m}$,  propagating from the left side ($x<0$) towards the barrier. Here, time $t$ is given in the units of $(H_0\gamma\mu_0)^{-1}$. The wave reflected from the barrier can be written as $R\,e^{i(-k_{m}x+\Omega t)}$, where $R$ is a complex amplitude. In the barrier region ($0<x<L$), the corresponding solution takes the form of the wave combination given in Eq.~(\ref{eq:gensol}), $\left(C_{1}e^{i k_{b}x}+C_{2}e^{-i k_{b}x}\right)e^{i\Omega t}$, where the evanescent solutions (with real exponents $\pm i k_b$) appear in the tunneling regime, i.e. for $1+\tilde{H}_{\rm a,m}<\Omega<1+\tilde{H}_{\rm a,b}$. In turn, the transmitted SW $T\,e^{i(k_{m}x+\Omega t)}$ is observed on the opposite side of the barrier (for $x>L$) with the complex amplitude $T$.

The transmissivity $T(\Omega,L)$ can be obtained by matching the solutions at the interfaces between the  barrier and matrix: (i) $(e^{i k_{m}x}+Re^{-i k_{m}x})e^{i\Omega t}$ with  $\left(C_{1}e^{i k_{b}x}+C_{2}e^{-i k_{b}x}\right)e^{i\Omega t}$ at $x=0$ and (ii)  $\left(C_{1}e^{i k_{b}x}+C_{2}e^{-i k_{b}x}\right)e^{i\Omega t}$ with $T e^{i(k_{m}x+\Omega t)}$ at $x=L$. To match the solutions we have to apply appropriate boundary conditions. The general formulation of the boundary conditions for exchange SWs should take into account possible change of exchange coupling at interfaces between the barrier and matrix.  This coupling affect the phase shift acquired by SW when it passes through the interface, and thus modifies the group delay $\tau_{gr}$. In our calculations we used Barnas-Mills boundary conditions (BMBC)~\cite{Barnas92,Mills92}, which include the modification of exchange  interaction at the interfaces both in the weak and strong coupling regime (see the Suplemantary Information for details). Using BMBC  one finds the transmissivity  $T(\Omega,L)$ in the form:
\begin{equation}
T(\Omega,L)=\frac{e^{-ik_{\rm m}L}}{\Delta_{c} \cos(k_{\rm b}L)+i\Delta_{s} \sin(k_{\rm b}L)},\label{eq:trans}
\end{equation}
where $\Delta_{s}(k_{\rm m},k_{\rm b})$ and $\Delta_{c}(k_{\rm m},k_{\rm b})$ are the rational expressions of the form depending on the boundary conditions on the interface between matrix and barrier  (see Sumplementary Information for details).

The transmissivity $T(\Omega,L)$ is one of the most important spectral characteristics of the system. Its magnitude $|T(\Omega,L)|$ gives the information about the energy density, which is transmitted through/over the barrier. For Eq.~(\ref{eq:trans}), we can write the following expression for $|T(\Omega,L)|$:
\begin{equation}
|T(\Omega,L)|=\big|\Delta_{\rm c} \cos(k_{\rm b}L)+i\Delta_{\rm s} \sin(k_{\rm b}L)\big|^{-1}.
\end{equation}
It is worth to notice that the transmissivity $T(\Omega,L)$ depends on  the barrier width $L$ only through the  factors: $\sin(k_{\rm b}L)$, $\cos(k_{\rm b}L)$ and $\exp(k_{\rm m}L)$, presented explicitly in Eq.~(\ref{eq:trans}).

The group delay $\tau_{gr}$ depends on the phase of the transmissivity function.
Following Ref.\citeonline{Winful06}, we find  $\tau_{gr}(\Omega,L)$ in the form:
\begin{equation}
\tau_{gr}(\Omega,L)=\frac{1}{\gamma\mu_{0}H_{0}}\frac{d}{d\Omega}\Big[{\rm Arg}\big(T\big)+k_{\rm m}L\Big].\label{eq:tau1}
\end{equation}
In the following, we will use the dimensionless group delay $\tilde{\tau}_{gr}(\Omega,L)$, defined as:
\begin{equation}
\tilde{\tau}_g(\Omega,L)=\gamma\mu_{0}H_{0}\tau_{gr}.
\end{equation}
The phase $\phi={\rm Arg}\big(T\big)+k_{\rm m}L$, gained by SW after tunneling (propagating) through (over) the barrier, consists of two terms. The term $k_{\rm m}L$ is a {\it geometrical} phase, which would be acquired by the SW on the distance $L$ (width of the barrier) in the absence of the barrier, i.e. propagating in  a homogeneous medium described by the material parameters of the matrix. The other term describes the phase shift resulting from the presence of the barrier. By referring to Eq.~(\ref{eq:trans}) we can notice that the phase $\phi$ can only be expressed by the argument of the denominator, $\Delta_{c} \cos(k_{\rm b}L)+i\Delta_{s} \sin(k_{\rm b}L)$. This allows writing  Eq.~(\ref{eq:tau1}) in a more explicit form:
\begin{equation}
\tilde{\tau}_{gr}(\Omega,L)=-\frac{d}{d\Omega}\Big[{\rm Arg}\big(\Delta_{\rm c} \cos(k_{\rm b}L)+i\Delta_{\rm s} \sin(k_{\rm b}L)\big)\Big].\label{eq:tau2}
\end{equation}

\subsection*{Hartman effect}

For tunneling SWs, the wave number $k_{\rm b}$ is purely imaginary:
\begin{equation}
k_{\rm b}=i\kappa_{\rm b},
\end{equation}
where $\kappa_{\rm b}$ is real.
Therefore, for a wide barrier ($L\gg1/\kappa_{\rm b}$), one can make the following simplifications in Eq.~(\ref{eq:tau2}): $\cos(k_{\rm b}L)=\cosh(\kappa_{\rm b}L)\approx\tfrac{1}{2}e^{\kappa_{\rm b}L}$, $\sin(k_{\rm b}L)=i\sinh(\kappa_{\rm b}L)\approx \tfrac{1}{2} i e^{\kappa_{\rm b}L}$. This brings us to the conclusion that the group delay $\tilde{\tau}_{gr}$ will saturate with increasing  barrier width ($L\rightarrow\infty$), which is the essence of the  \textit{Hartman effect}. In this limit, the group delay becomes independent on the barrier width:
\begin{equation}
\tilde{\tau}_{gr}(\Omega)\underset{L\rightarrow\infty}{=}-\frac{d}{d\Omega}\Big[{\rm Arg}(\Delta_{\rm c} -\Delta_{\rm s})\Big].\label{eq:tau4}
\end{equation}
It can be proved that the group delay in tunnelling regime is always positive. Thus, the controversies related to Hartman effect (including the discussion about the violation of causality) have nothing to do with negative group delay. In the limit $L\rightarrow\infty$ the group delay diverges at the ranges of tunnelling regime, for $\Omega=1+\tilde{H}_{\rm a,m}$ and $\Omega=1+\tilde{H}_{\rm a,b}$.

One of the main obstacles making the observation of the HE difficult is the low magnitude of the transmissivity $|T(\Omega,L)|$ in the tunneling regime, which decays exponentially with increasing  barrier width $L$ for $L\gg1/k_{\rm b}$, where one finds:
\begin{equation}
|T(\Omega,L)|\underset{L\gg1/\kappa_{\rm b}}{=}e^{-\kappa_{\rm b}L}\big|\Delta_{\rm c}-\Delta_{\rm s})\big|^{-1}.
\end{equation}
Therefore, it is useful to define the so-called figure-of-merit (FOM) for the HE:
\begin{equation}
{\rm FOM}=\frac{|T|}{\tilde{\tau}_{gr}}.
\end{equation}
 The high value of the FOM points out the parameters of the model for which the short group delay coincides with relatively high magnitude of the transmissivity.

 The HE can be also observed for reflected waves, where the group delay is defined as\cite{Winful06}:
\begin{equation}
\tau'_{gr}(\Omega,L)=\frac{1}{\gamma\mu_{0}H_{0}}\frac{d}{d\Omega}\Big[{\rm Arg}\big(R\big)\Big].\label{eq:tauR}
\end{equation}
The complex coefficient $R$ is the reflectivity (see the discussion the beginning of this section). For a symmetric barrier, the group delays for transmitted waves $\tau_{gr}$ and for reflected waves $\tau'_{gr}$ are equal\cite{Flack88,Winful03c}. This also means that for the barrier  characterized by a symmetric shape of the effective anisotropy field: $H_{\rm a}(L/2+x)=H_{\rm a}(L/2-x)$ and identical boundary conditions at $x=0,L$, the saturated values of the group delay  for transmitted and reflected waves will be also equal.

\section*{Numerical results}

\begin{figure}[t!]
\centering
\centering
\includegraphics[width=\columnwidth]{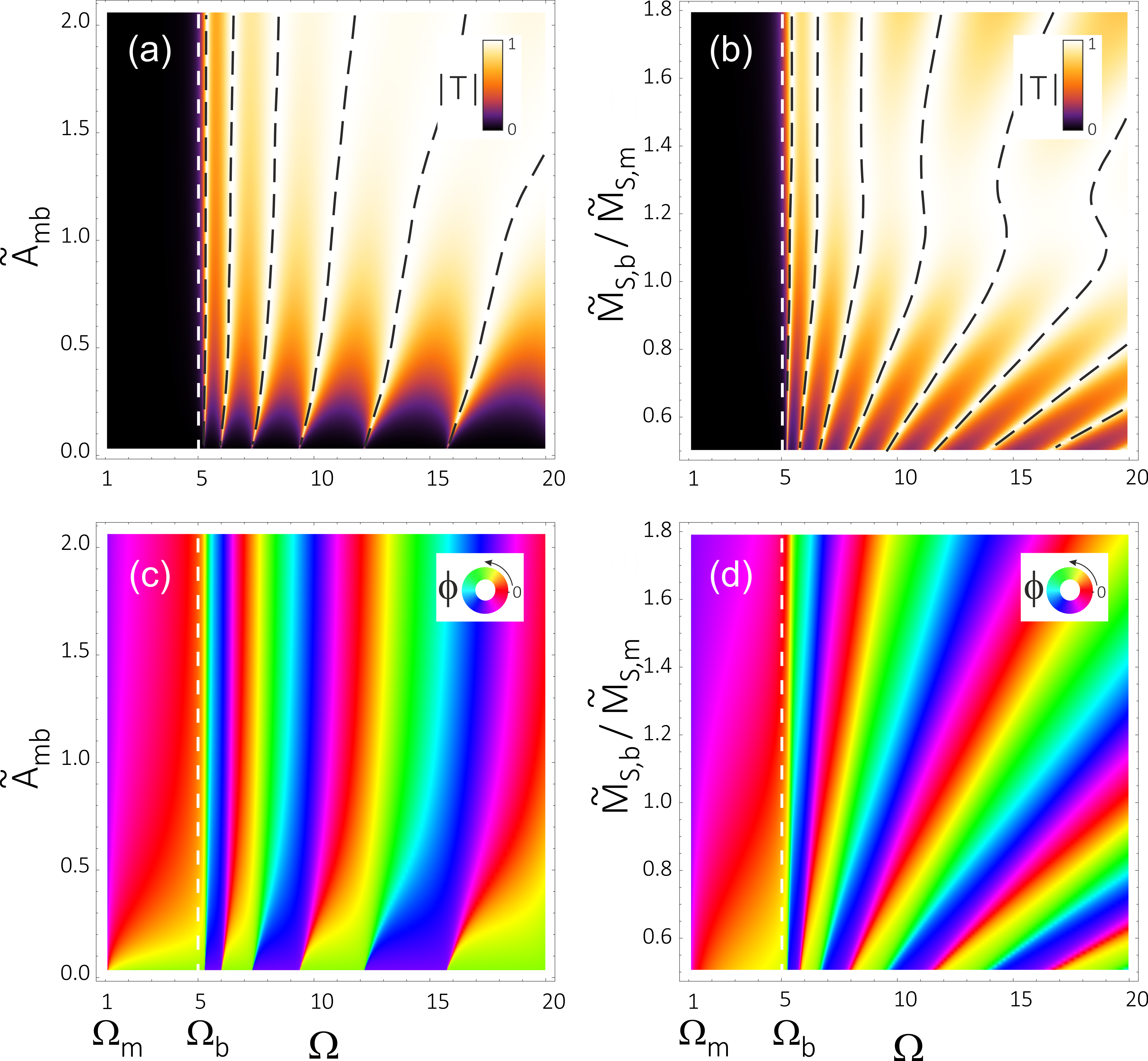}
\caption{Absolute value of the transmissivity $|T|$ (a,b) and the phase $\phi$ gained in the transmission (c,d) for the anisotropy barrier of height $\tilde{H}_{\rm a,b}+1=5$ and width $L=5$, separated from the matrix by an interface layer of width $t=0.25$ (widths are given in the units of $\lambda_{ex}$). Both $|T|$ and $\phi$ are presented  as a function of the spin wave frequency $\Omega$ and material parameters: strength of the interface exchange coupling, $\tilde{A}_{\rm mb}=\lambda_{\rm{ex,mb}}^2\tilde{M}_{\rm{S,mb}}^2$ (a,c) and the magnetization contrast between the barrier and matrix, $\tilde{M}_{\rm S,b}/\tilde{M}_{\rm S,m}$ (b,d). Black dashed lines in (a,b) mark the maxima ($|T|=1$) of the transmissivity. The frequencies $\Omega_{\rm m}=\tilde{H}_{\rm a,m}+1=1$ and $\Omega_{\rm b}=\tilde{H}_{\rm a,b}+1=5$ denote the minimal frequency for the propagating exchange SWs in homogeneous materials of the matrix and barrier, respectively. The later one is marked additionally by vertical white dashed line. The calculations have been done for the same values of exchange length $\lambda_{\rm ex}=1$ in the barrier and in the matrix. The width of barrier and the exchange length are measured in the same a.u. of length.}
\label{fig:Fig2}
\end{figure}

Now we present numerical results obtained for the system under consideration. In Fig.~\ref{fig:Fig2} we show the absolute value of the transmissivity $|T|$ (a,b) and the phase $\phi={\rm Arg} (T)+k_{\rm m}L$ gained by SW transmitted through (or over) the magnetic barrier (b,d) -- both as a function of the frequency $\Omega$ and two selected model parameters (one of interface and another one of bulk character).
More specifically, we show there the impact of interface exchange coupling (defined as $\tilde{A}_{\rm mb}=\lambda_{\rm{ex,mb}}^2 M_{\rm{S,mb}}^2/H_0^2$ -- see Fig.\ref{fig:Fig1} and Supplementary Information) (a,c),  and the influence of a  contrast between the bulk  saturation magnetization of the barrier, $M_{\rm{S,b}}$, and of the matrix, $M_{\rm{S,m}}$ (c,d). The results have been  obtained with the use of general BMBC. These conditions, however, comprise other types of the boundary conditions  considered here (see Supplementary Information).

Transmission in the  tunneling regime (under barrier), i.e. for frequencies $\Omega<\Omega_{\rm b} =\tilde{H}_{\rm a,b}+1$), is very small, but it increases rapidly for $\Omega$ approaching  $\Omega_{\rm b}$ ($\Omega_{\rm b}=5$ in Fig.~\ref{fig:Fig2}). This increase is less rapid
for stronger exchange coupling between the barrier and matrix (see Fig.~\ref{fig:Fig2}a), and for the saturation magnetization in the barrier larger than that in the matrix region (see Fig.~\ref{fig:Fig2}b).
\begin{figure}[t!]
\centering
\includegraphics[width=\columnwidth]{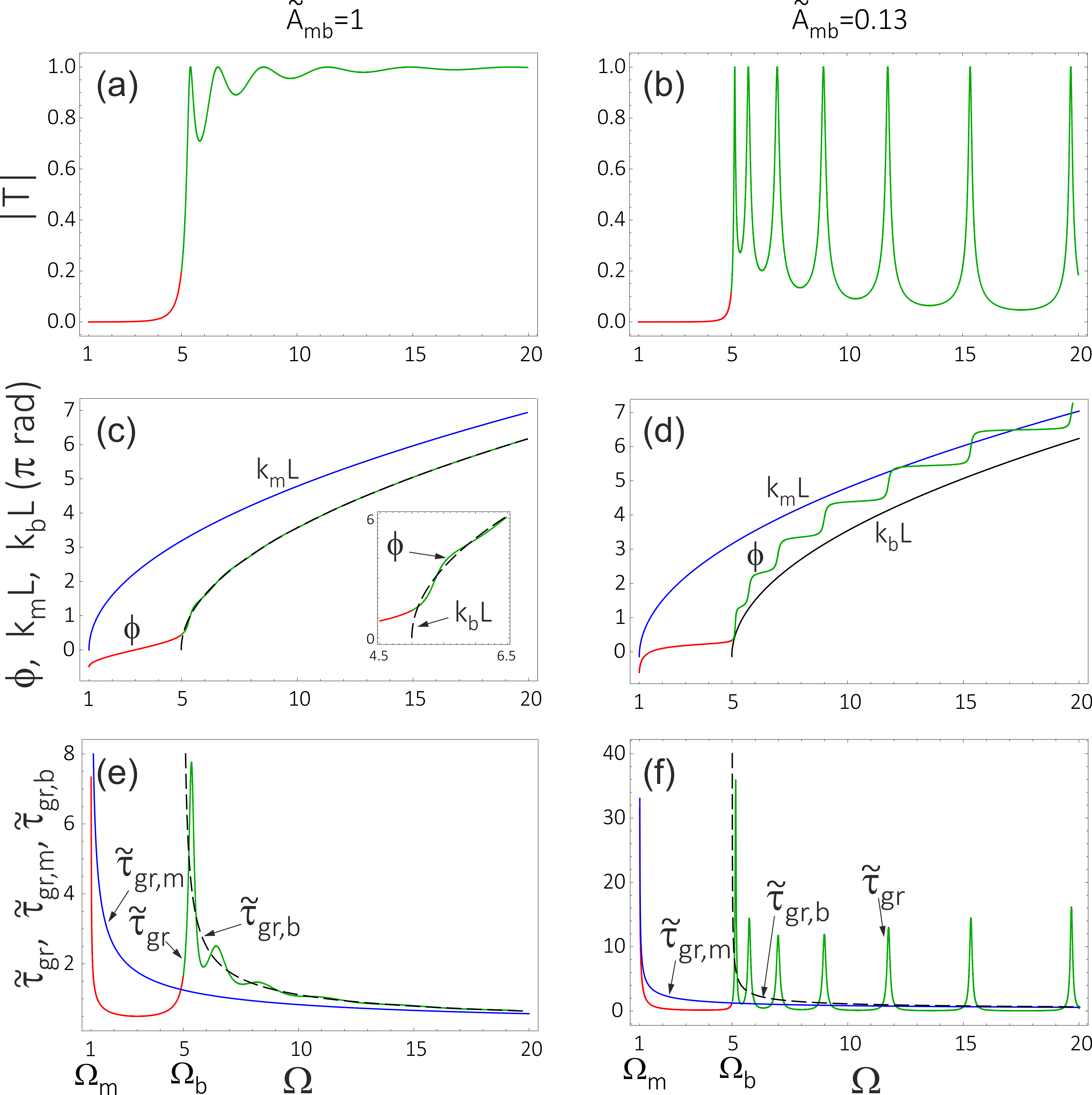}
\caption{Absolute value of the transmissivity $|T|$ (a,b), the phase $\phi$ gained in the transmission (c,d), and the corresponding group delay (e,f) shown as a function of the spin wave frequency $\Omega$. These parameters are shown  for the anisotropy barrier of the same parameters as in Fig.~\ref{fig:Fig2}, and  for two different strengths of the exchange coupling  $\tilde{A}_{\rm mb}$  between the barrier and matrix. The first column (a,c,e) presents the results for an intermediate strength of the coupling (which can be reduced to the case where the natural boundary conditions are applicable -- see Supplementary Information), while the second column (b,d,f) shows the results in the regime of week coupling (for which the Hoffmann boundary conditions can be used -- see Suplementary Information). Different colors correspond to: tunneling through the barrier (red), propagation over the barrier (green), propagation in the homogeneous material: matrix (blue) or barrier (black). The calculations have been done for the same values of $\tilde{M}_{\rm S}=1$ and $\lambda_{\rm ex}=1$ in the barrier and in the matrix.
}
\label{fig:Fig3}
\end{figure}

For SWs propagating at frequencies $\Omega>\Omega_{\rm b}$ (over barrier transmission), the transmissivity $T$ oscillates with increasing $\Omega$, see the maxima (resonances) indicated by the dashed lines in Fig.~\ref{fig:Fig2}(a,b), for which $|T(\Omega)|=1$. For $\Omega>\Omega_{\rm b}$, the modulus of transmissivity, $|T|$, reveals sharp peaks in the regime of  weak exchange coupling between the matrix and barrier (region of small interface exchange parameter in Fig.~\ref{fig:Fig2}a). The weakest oscillations of $|T|$ are observed for moderate values of the interface coupling ($\tilde{A}_{\rm mb}=1$ in Fig.~\ref{fig:Fig2}a), corresponding to the natural boundary conditions (NBC) -- see Suplementary Information . The oscillation amplitude of $|T|$ increases again  with increasing interface exchange parameter, which leads to deeper minima between the resonances. This behavior can be  even more clearly seen in Fig.~\ref{fig:Fig3}(a,b), where $|T|$ is shown as a function of $\Omega$ for two selected values of the interface exchange parameter, corresponding to the natural boundary conditions (Fig.~\ref{fig:Fig3}a) and a weak interface coupling (Fig.~\ref{fig:Fig3}b).

The above described properties of the transmissivity are similar to those observed for tunneling  of other waves existing in  nature, including tunneling of particles in quantum mechanics~\cite{Winful06}. The exact correspondence to quantum mechanical tunneling can be strictly shown for the NBC, when we neglect the contrast of $M_{\rm{S}}$ between the matrix and barrier (see Fig.~\ref{fig:Fig3}a).

The phase acquired in the transmission through (over) the barrier, $\phi={\rm Arg}(T)+k_{\rm{m}}L$, is an important parameter describing the dynamical properties of wave propagation. Due to fast changes of $\phi$ in the frequency domain, we observe large values of the group delay  $\tau_{gr}$ (see Eqs.~\ref{eq:tau1},~\ref{eq:trans}). From Figs.~\ref{fig:Fig2}(c,d) one can also notice that the phase $\phi$ grows (circulates) monotonously with the frequency $\Omega$. Therefore, the group delay is positive as one might expect. The changes of $\phi$ (and also  of $\tau_{gr}$) vary, however, in the frequency domain. In a homogeneous system corresponding, e.g., to the matrix, the phase $\phi_{\rm m}$ gained on the distance $L$ is proportional to the wave number $k_{\rm m}$, $\phi_{\rm m}=L k_{\rm m}$. Due to a quadratic dispersion relation of the exchange SWs, the phase $\phi_{\rm m}$  changes in the frequency domain as $\sqrt{\Omega}$, shifted by a constant value resulting from the  static effective field. The magnetic barrier introduces an additional term, ${\rm Arg} (T)$, to the phase $\phi$. This term reflects two features of the barrier, which influence the phase of SWs: (i) the difference in effective anisotropy fields and the contrast of material parameters (saturation magnetization and exchange length), which affect the wave number (see Eqs.~\ref{eq:gensol},\ref{eq:wavenum}); (ii) the strength of exchange coupling at the interface between the barrier and matrix included in the boundary conditions (see Supplementary Information), which determines the jump of the phase at these interfaces. For moderate coupling of the barrier and matrix, the correction ${\rm Arg} (T)$ makes the $\phi(\Omega)$  relation similar to that in the homogeneous system made of the material used to create the  barrier, $\phi\approx k_{\rm b}L$ (see Fig.\ref{fig:Fig3}c), with some hardly noticeable deviation close to $\Omega_{\rm b}$ and at the frequencies corresponding to the transmission resonances, where $|T(\Omega)|=1$ (see the inset in Fig.~\ref{fig:Fig3}c). These oscillations in the slope of $\phi(\Omega)$ are responsible for the peaks in the group delay $\tau_{gr}$, clearly seen in Fig.~\ref{fig:Fig3}e. Note that for the homogeneous systems  $\tau_{gr}(\Omega)$ decays monotonously with increasing $\Omega$ (see the solid blue and dashed black lines in Fig.~\ref{fig:Fig3}e). The oscillations in ${\rm Arg} (T)$ are related to the transmissivity resonances, $|T|=1$.  The phase increases approximately by $\pi$ between two successive resonances (this rule is strict for strong interface exchange coupling). Due to a quadratic dispersion relation ($\Omega \propto k_{\alpha}^{2}=(\phi_{\alpha}/L)^2$), the distance between successive resonances and peaks of the group delay increases.

The impact of interface exchange coupling and contrast of magnetization (between the barrier and matrix) on the group delay can be deduced from Fig.~\ref{fig:Fig2}(c,d). The following conclusions can be drawn for the propagation regime ($\Omega>\Omega_{\rm b}$): For weaker interface exchange coupling $\tilde{A}_{\rm mb}$ and  $\tilde{M}_{\rm S}$ in the barrier lower than  $\tilde{M}_{\rm S}$ in the matrix one finds, (i)  the phase $\phi$ changes more rapidly in the frequency domain and therefore the peaks in $\tau_{gr}$ are expected to be higher, and (ii) there are more phase oscillations and more peaks in $\tau_{gr}$ per frequency unit. These changes can be attributed to the reduction of spin wave pinning for the weaker interface exchange coupling $\tilde{A}_{\rm mb}$ and extension of the wavelength in the barrier for lower $\tilde{M}_{\rm S}$ inside the barrier.

From Figs.~\ref{fig:Fig2}(c,d) one can conclude that in the tunneling regime, $\Omega_{\rm m}<\Omega<\Omega_{\rm b}$, the phase changes more rapidly just above the lowest allowed  frequency of propagating SWs in the matrix, $\Omega_{\rm m}=\tilde{H}_{\rm a,m}+1$, and just below the frequency $\Omega_{\rm b}$ (determining the threshold between tunneling and propagaton regime). At these frequencies, the slope of the $\phi(\Omega)$ dependence is infinite for homogeneous materials of both the matrix and the barrier. This results in infinite values of the group delay $\tau_{gr}$ for homogeneous materials (at the mentioned frequencies), which is supposed to influence the value of $\tau_{gr}$ for the system composed of the barrier embedded in the matrix. It is also worth to notice that the phase shift $\phi$  (between the transmitted and incoming wave) is surprisingly negative for the very low frequencies (close to the $\Omega_{\rm m}$) and then,  it increases for larger frequencies and reaches positive values close to $\Omega_{\rm b}$.

Figure~\ref{fig:Fig3} shows the  modulus of the transmissivity $|T(\Omega)|$ (a,b), the phase $\phi(\Omega)$ gained by SW after transmission (c,d), and the corresponding group delay $\tau_{gr}(\Omega)$ (e,f). The numerical results on $\phi(\Omega)$ and $\tau_{gr}(\Omega)$ for the barrier embedded in the matrix are supplemented by the phase of SWs propagating in the homogeneous material of the barrier and matrix,  acquired at the distance equal to the width of the barrier, as well as with the plots of the corresponding  group delays. These plots have  been obtained for the structure without contrast of bulk magnetic parameters ($\tilde{M}_{\rm S}$ and $\lambda_{\rm ex}$). Using the BMBC we analyzed two special cases: intermediate exchange coupling at the interface, corresponding to the NBC, and weak interface coupling, where the Hoffman boundary conditions (HBC) are applicable\cite{Hoffmann70a} -- see Supplementary Information. The plots in Fig.~\ref{fig:Fig3}(a,c) are strictly counterparts of the corresponding plots for electronic waves, reported e.g in Ref.\citeonline{Winful06}. The plots presented in Fig.~\ref{fig:Fig3} give more detailed insight into the transmission properties of SWs tunnelling (propagating) through (over) the magnetic barrier. The  interesting effect is observed in Fig.~\ref{fig:Fig3}(b). For small value of $A_{\rm mb}$, we can see the resonate  transmission of spin waves. It means that in such conditions the  barrier region is almost decoupled from the matrix and the transmission peaks are supposed to be sharp even for the spin waves of high frequency propagating over the barrier. For lower frequencies, we observed that the group delay of tunnelling spin waves $\tilde{\tau}_{\rm gr}$ (red line in Fig.~\ref{fig:Fig3} (e,f)) is shorter than in uniform space $\tilde{\tau}_{\rm gr,m}$ (blue line in Fig.~\ref{fig:Fig3} (e,f)) by the cost of reduction of transitivity $|T|$ (compare the red lines in Fig.~\ref{fig:Fig3} (a,b) ). This difference increases for weaker coupling $A_{\rm mb}$ (compare the red lines in Fig.~\ref{fig:Fig3} (e,f) ).

\begin{figure}[t!]
\centering
\includegraphics[width=\columnwidth]{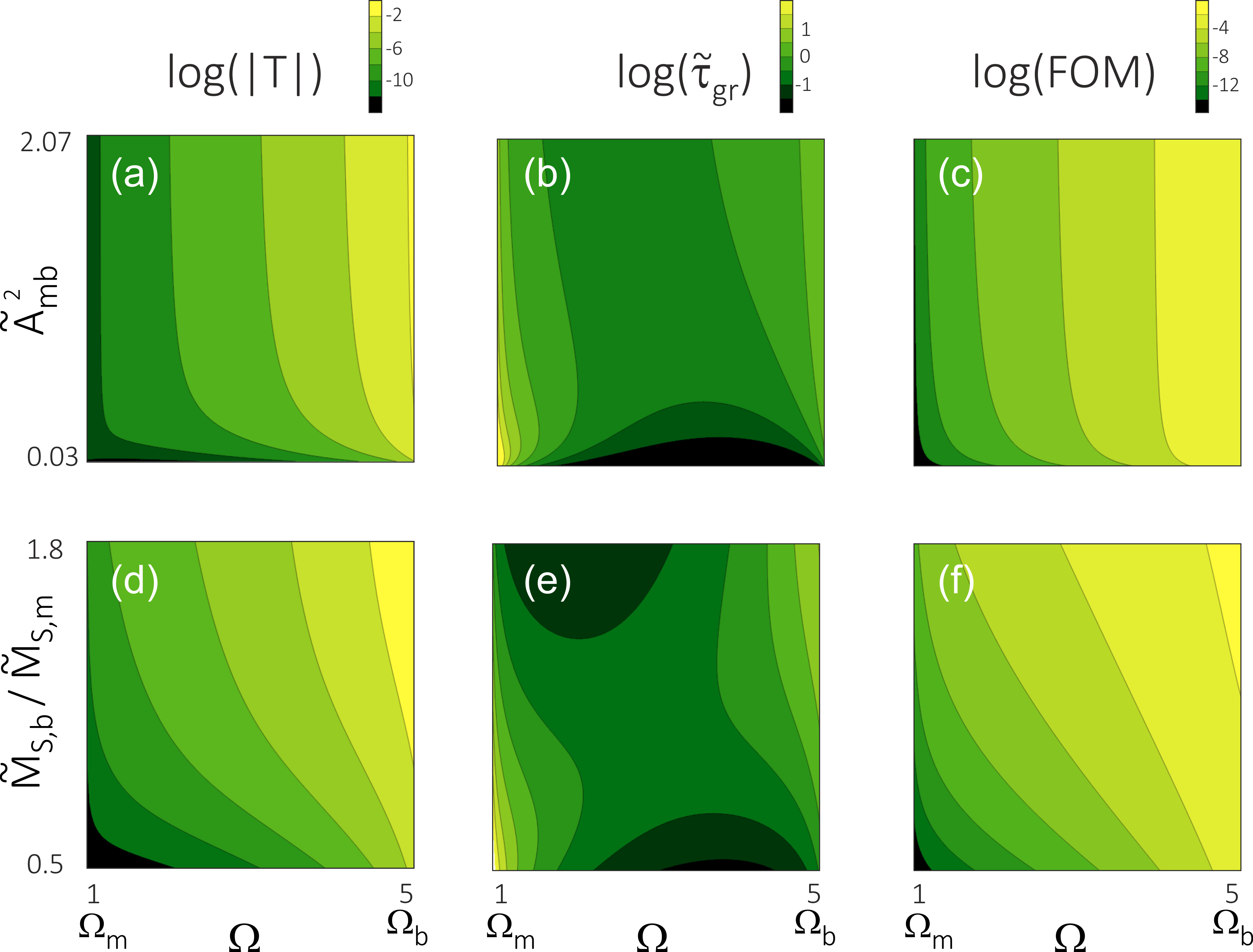}
\caption{Logarithm of the transmissivity modulus (a,d), group delay (b,e), and FOM (c,f) in the tunneling regime, $\Omega_{\rm m} < \Omega < \Omega_{\rm b}$. All parameters are presented as a function of spin wave frequency $\Omega$ and material parameters: strength of the interface exchange coupling $\tilde{A}_{\rm mb}$ (a-c) and the magnetization contrast between the barrier and matrix, $\tilde{M}_{\rm S,b}/\tilde{M}_{\rm S,m}$ (d-f). The calculations have been done for the same values of the exchange length $\lambda_{\rm ex}=1$ in the barrier and in the matrix. The parameters of the barrier are the same as the ones used in Fig.~\ref{fig:Fig2}.}
\label{fig:Fig4}
\end{figure}

To discuss the HE one needs to analyze in detail the properties of the considered system in the tunneling regime. In  Fig.~\ref{fig:Fig4} we show the absolute value of the   transmissivity moduls $|T|$ (a,d), the group delay $\tilde{\tau}_{\rm gr}$ (b,e), and the corresponding FOM (c,f) -- all as a function of frequency $\Omega$ and  material parameters: interface exchange coupling (a-c) and contrast of saturation magnetization between the barrier and matrix (d-f).  All the quantities, that is $|T|$, $\tilde{\tau}_{\rm gr}$ and FOM, are shown in the logarithmic scale. The yellow regions in this figure correspond to large values of the corresponding parameter.

The observation of HE requires optimally large  amplitude of the tunneling SWs. This means that one should consider spin wave packets in the frequency range not very distant from the threshold frequency $\Omega_{\rm b}$ corresponding to the top of the magnetic barrier. The other requirement is a relatively small value of the group delay, which can allow the detection of SWs after passing the magnetic barrier of a certain width in the presence of damping. The last column in Fig.~\ref{fig:Fig4} (see (c,f))  shows the FOM defined as the ratio of tunnelling amplitude and group delay. The yellow regions indicate the range of parameters which are the most suitable for experimental observation of the HE.

The modulus of the transmissivity $|T|$ decays exponentially with decreasing frequency. Therefore, for practical application, only the higher range of frequencies, close to $\Omega_{\rm b}$, is of some interest. The increase of the interface exchange coupling or saturation magnetization in the barrier can slightly extend  this range towards lower frequencies. The group delay, $\tilde{\tau}_{\rm gr}$,  reaches the lowest values for intermediate frequencies, between the lowest frequencies for propagating modes in the matrix (here $\Omega_{\rm m}=1$) and in the barrier (here $\Omega_{\rm b}=5$). The further lowering of $\tilde{\tau}_{\rm gr}$ could be achieved by reducing the interface exchange coupling or saturation magnetization in the barrier (with respect to that in the matrix) -- see Fig.~\ref{fig:Fig4}(b,e). Unfortunately, this change simultaneously leads to a decrease of $|T|$. The better strategy is thus to increase the saturation magnetization contrast by taking larger values of $\tilde{M}_{\rm S}$ in the barrier (see Fig.~\ref{fig:Fig4}e). However, the changes of $\tilde{\tau}_{\rm gr}$ with frequency $\Omega$ are not so large as the changes in the transmissivity modulus  $|T|$. Therefore, the decisive factor for increasing the FOM, which makes the observation of the HE possible, is the optimization of the modulus of the transmissivity by selection of the frequency range slightly below the threshold value $\Omega_{\rm b}$, and selection of appropriate values of material parameters (e.g. by the increase of $\tilde{M}_{\rm S}$ in the barrier region with respect to that in the matrix).

\begin{figure}[t!]
\centering
\includegraphics[width=\columnwidth]{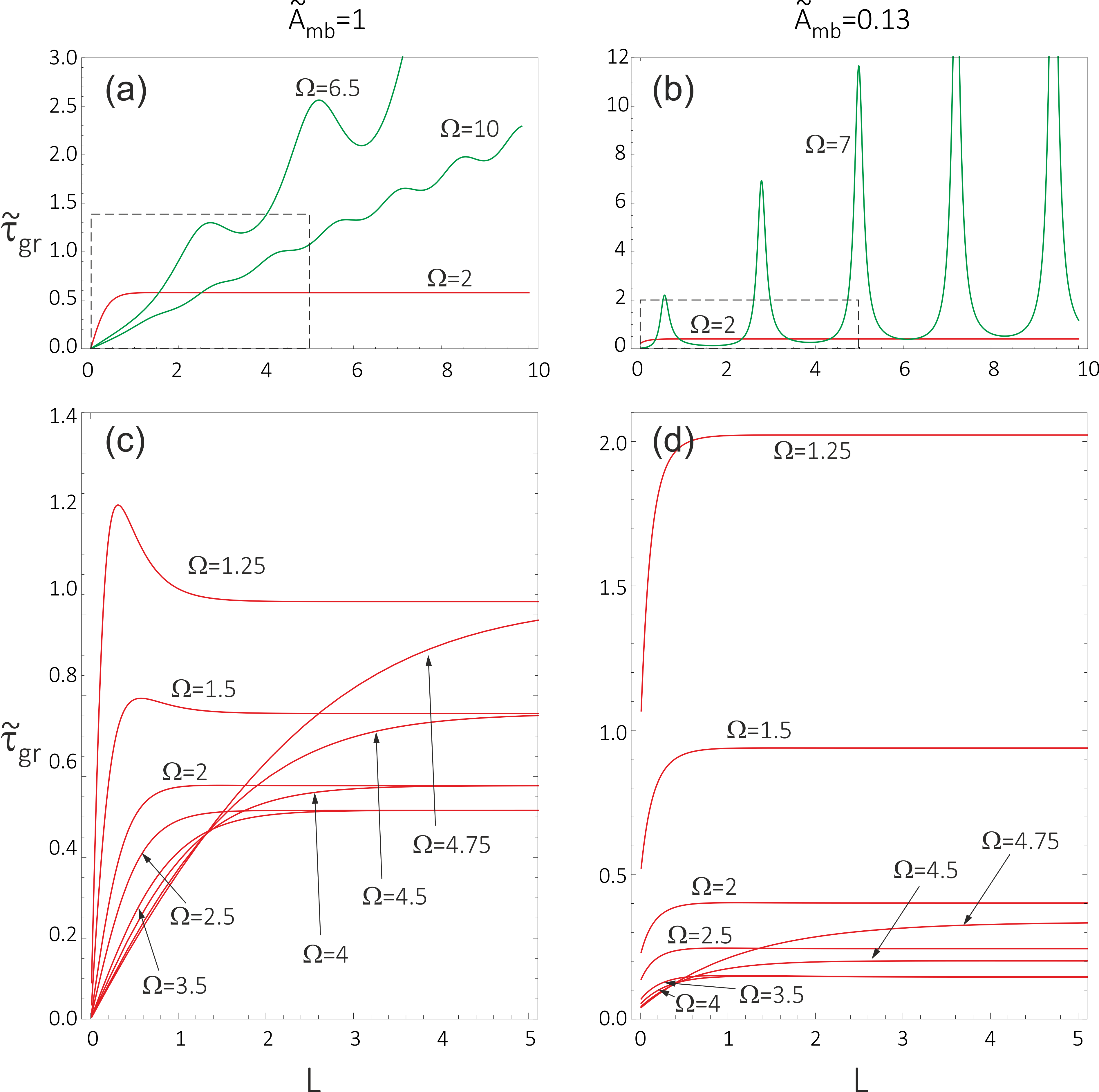}
\caption{Dependence of the group delay $\tau_{gr}$ on the barrier width $L$ for the selected frequencies $\Omega$  and for two different strengths of interface exchange coupling $\tilde{A}_{\rm mb}$ between the barrier and matrix. The first column (a,c) presents the results for intermediate strength of the coupling (which can be reduced to the case where the natural boundary conditions are applicable). The second column (b,d) shows the results in the regime of week coupling (for which the Hoffmann boundary conditions can be used). The saturation of the group delay, being the signature of HE, appears in the tunneling regime ($\Omega_{\rm m}<\Omega<\Omega_{\rm b}$) and is shown in more details in (c,d). The calculations were performed for the same values of $\tilde{M}_{\rm S}=1$ and $A_{\rm ex}=1$ in the barrier and matrix. The parameters of the barrier are the same as the ones used in Fig.~\ref{fig:Fig2}.}
\label{fig:Fig5}
\end{figure}
\begin{figure}[t!]
\centering
\includegraphics[width=\columnwidth]{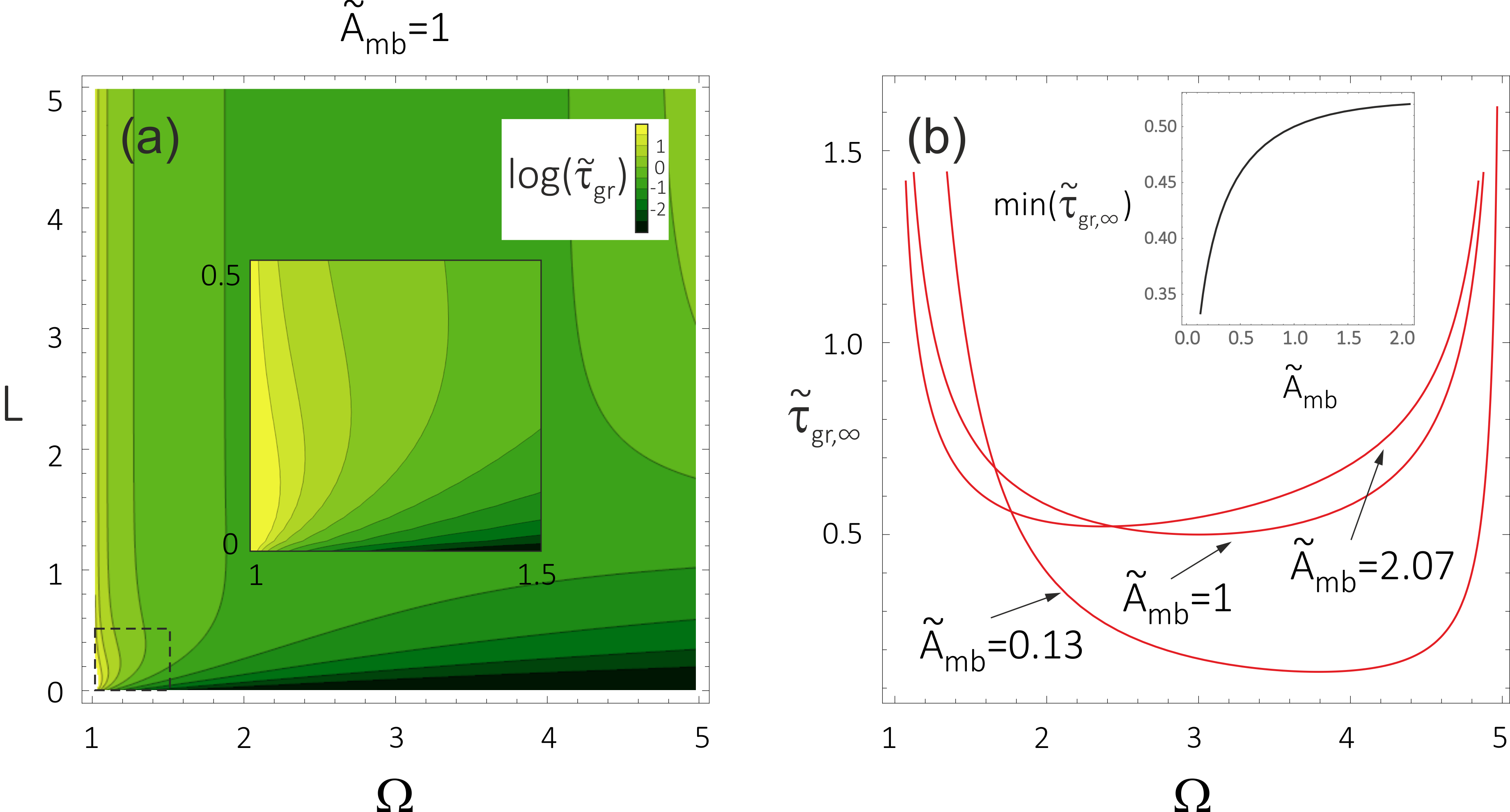}
\caption{(a) Group delay as a function of the barrier thickness and frequency for the natural boundary conditions (see Supplementary Information) applied at the interface between the barrier and matrix. The peculiarities related to the non-monotonous dependence of $\tau_{gr}(L)$  are zoomed in the inset. (b) Frequency dependence of the saturation value of the group delay,  $\tau_{gr}(L\to\infty)$, for the indicated values of the exchange coupling between the barrier and matrix $\tilde{A}_{\rm mb}$. The calculations have been done for the same values of $\tilde{M}_{\rm S}=1$ and $\lambda_{\rm ex}=1$ in the barrier and matrix. The inset in (b) presents the dependence of minimal $\tau_{gr}(L\to\infty)$ on $\tilde{A}_{\rm mb}$. The parameters of the barrier are the same as those used in Fig.~\ref{fig:Fig2}.}
\label{fig:Fig6}
\end{figure}

Now we present the results which demonstrate the HE for the exchange-dominated SWs.  Figure~\ref{fig:Fig5} shows the saturation of the group delay with increasing barrier width in the tunneling regime (red curves), which can be considered as manifestation of the HE. The presented results (Fig.~\ref{fig:Fig5}) have been obtained for the BMBC, which in the regime of intermediate interface exchange coupling (a,c)  and weak exchange coupling (b,d)  reduce to the NBC and the HBC, respectively.
Let us analyze these results in more details. First, we note that the group delay for the {\it under-barrier} tunneling behaves in a different way than that for the {\it over-barrier} propagation. In the former case, the group delay at small thicknesses $L$ is larger than the time which SW needs to traverse the distance $L$  in the free space (no barrier). In the free space, i.e. in the homogeneous medium made of the matrix material, the group delay increases linearly with the distance, $\tilde{\tau}_{gr,\alpha}=L\frac{d k_{\rm,\alpha}}{d \Omega}=L v_{gr,\alpha}$, where $v_{gr,\alpha}$ is the group velocity in the homogeneous material. For larger values of $L$, the group delay increases more slowly than in the case of free motion. Moreover, the group delay saturates with increasing $L$ (see the red lines in Fig.~\ref{fig:Fig5}(a,b)).
In turn, the group delay for the {\it over-barrier} propagation reveals oscillations with increasing $L$, but overall it increases linearly with increasing $L$ (see the green curves in Fig.~\ref{fig:Fig5}a plotted for the NBC, which correspond to an intermediate strength of the interface exchange coupling). The oscillations in the group delay can be significantly stronger in the regime of weak interface exchange coupling, see the green curves in Fig.~\ref{fig:Fig5}b, where the HBC can be applied. The observed peaks in $\tilde{\tau}_{gr}$ (and also those in $|T|$) are related to the resonant tunneling, which can be achieved by the selection of frequency/wavelength  or by adjusting the width of the barrier. It is worth to note that even in the regime of weak interface coupling (see the green curves in Fig.~\ref{fig:Fig3}(b,f)) the linear growth with $L$ is observed both for the maxima of $\tilde{\tau}_{gr}$ peaks and for the minima between them.

In the tunneling regime ({\it under-barrier} transmission), the group delay $\tilde{\tau}_{gr}(L)$ saturates with increasing $L$ (see Fig.~\ref{fig:Fig5}(c,d)). The saturation with increasing $L$ is slower for the higher frequencies which are closer to the threshold value $\Omega_{\rm b}$. This property makes the observation of the HE difficult because it requires to use wide barriers. Note, that this frequency range is characterized by high values of the FOM, which is beneficial for spin wave transmission. It is also worth to note that for higher, or even intermediate strengths of the interface exchange coupling, the dependence of the group delay on the barrier width is non-monotonous for the lowest frequencies (see the curves for $\Omega=1.25,1.5$ in Fig.~\ref{fig:Fig5}c.

There is one interesting feature of the group delay curves shown in Fig.~\ref{fig:Fig5}(c,d). Namely, in Fig.~\ref{fig:Fig5}(c) the group delay  vanishes in the limit $L\to 0$, while in Fig.~\ref{fig:Fig5}(d) a small nonzero values of the group delay remains when $L=0$. This follows from the fact that although $L$ is reduced to zero, the modified exchange coupling at the barrier/matrix boundaries ($\tilde{A}_{\rm mb}=0.13$)  remain in Fig.~\ref{fig:Fig5}(d) when $L$ is reduced to zero. To achieve a uniform system in the limit of $L=0$ one should also restore the full coupling limit.

It is clear from Fig.~\ref{fig:Fig5}(c,d) that the saturation level of the  group delay changes with the frequency, displaying a minimum in the tunnelling range $\Omega_{\rm m}<\Omega<\Omega_{\rm b}$. This is shown explicitly in Fig.~\ref{fig:Fig6}. From Fig.~\ref{fig:Fig6}(a) follows that for the barrier thickness $L=5$ (assumed earlier  for calculations presented in Figs.~\ref{fig:Fig2},~\ref{fig:Fig3},~\ref{fig:Fig4}), the group delay is almost saturated. Figure~\ref{fig:Fig6}(a) was obtained for the NBC in the absence of saturation magnetization contrast (like Fig.~\ref{fig:Fig3}(a,c,e) and Fig.~\ref{fig:Fig5}(b,c)). We can trace here in detail the effect of reducing the slope of $\tilde{\tau}_{gr}(L)$ with increasing $\Omega$ and the non-monotonic character of the $\tilde{\tau}_{gr}(L)$ relation for the lowest frequencies (shown in the inset).

The changes of saturation level $\tilde{\tau}_{\rm gr,\infty}$ with the frequency $\Omega$ are shown even more clearly in Fig.~\ref{fig:Fig6}(b) for few selected values of the exchange coupling between matrix and barrier: $A_{\rm mb}$. The minimum of this dependence becomes deeper with decreasing $A_{\rm mb}$ (see inset in Fig.~\ref{fig:Fig6}(b)), and is shifted towards higher frequencies. The shorter group delay is beneficial for observation of the HE. The better strategy to increase the FOM for the HE is to select the central frequency of tunnelling wave package close to the minimum of the dependence: $\tilde{\tau}_{\rm gr,\infty}(\Omega)$  than to reduce the coupling: $A_{\rm mb}$. Because the later approach will result in significant the decrease of the $|T|$. 

By reducing of $\tilde{\tau}_{\rm gr,\infty}$ we can potentially gain, for narrower barriers, the condition: $\tilde{\tau}_{\rm gr}<\tilde{\tau}_{\rm gr,m}$ (delay time for tunneling is shorter than delay in uniform space). This condition is more useful for experimental search for the signatures of the HE than looking for the saturation of group delay with the increase of the barrier width. 

\begin{figure}[t!]
\centering
\includegraphics[width=\columnwidth]{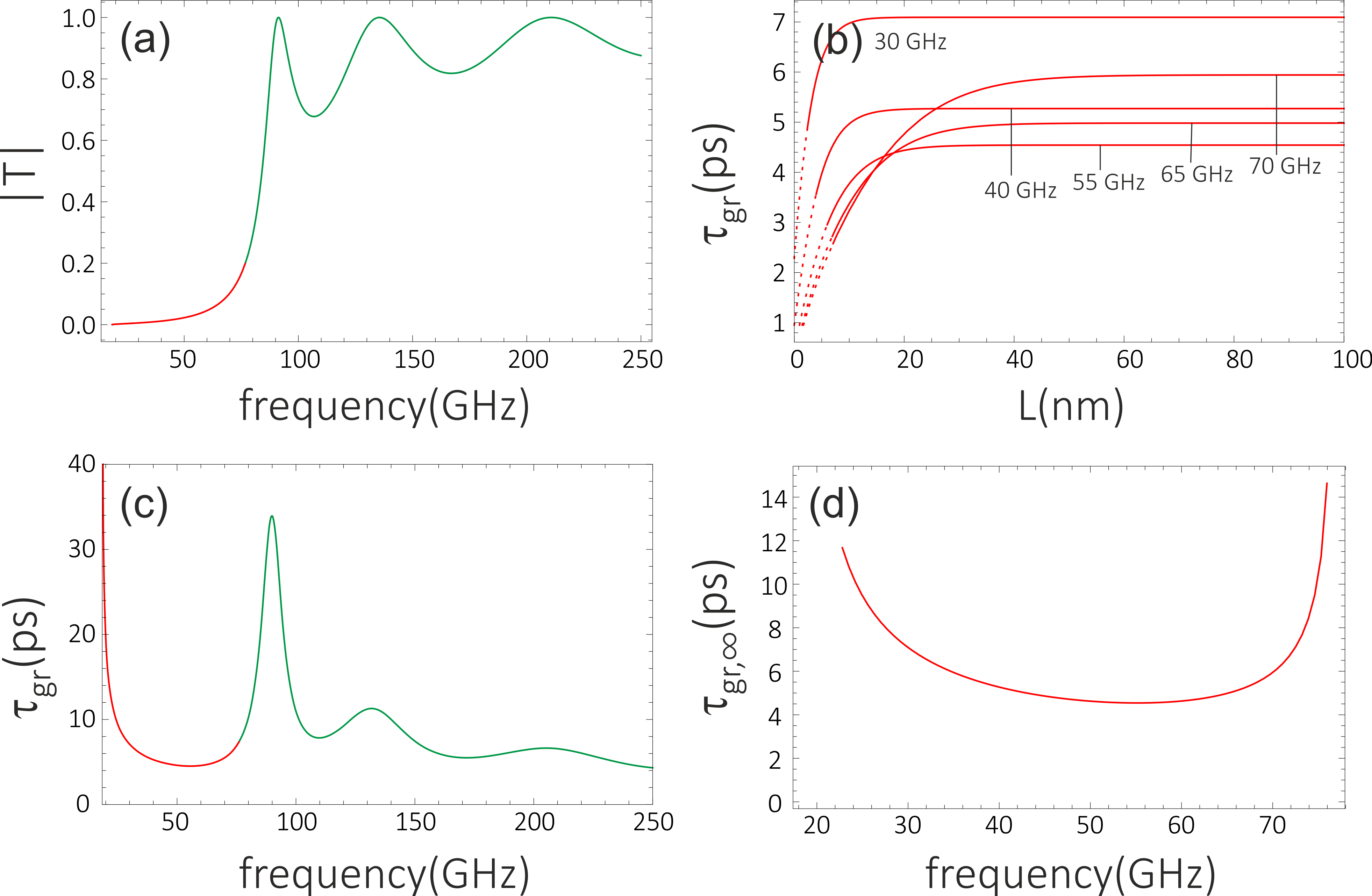}
\caption{Modulus of the transmissivity
$|T|$ (a) and the corresponding group delay $\tau_{gr}$ (c) for the CoFeB layer of thickness $t_{\rm CoFeB,m}$=1.3 nm and $t_{\rm CoFeB,b}$=1.0 nm in the matrix and the barrier region, respectively. For $|T|$ and $\tau_{gr}$  we assumed  the width of the barrier $L$=30 nm. Thickness of the interface between the matrix and the barrier was assumed as $t_{\rm tm}$=4 nm. The following material parameters were assumed: $M_{\rm S,m}=1.2\times10^6$ A/m, $M_{\rm S,b}=0.8\times10^6$ A/m, $A_{\rm m}=27\times10^{12}$ J/m,  $A_{\rm b}=20\times10^{12}$ J/m, $K_{\rm i}=1.3\times10^{-3}$ ${\rm J/m^3}$. The calculations were performed for external field $\mu_0H_0$=0.5T. Saturation of the group delay $\tau_{gr}$, observed in the tunneling regime, is shown in (b). The change of the saturation value of $\tau_{gr,\infty}$ as a function of frequency (d) shows the minimum around 55 GHz.
}
\label{fig:Fig7}
\end{figure}

To check how the model described above refers to real systems (see Fig.\ref{fig:Fig1}), we performed numerical calculations for a thin layer of CoFeB, which is slightly thinner in the barrier region ($t_{\rm l,b}=t_{\rm CoFeB,b}=1.0$ nm) than in the matrix area ($t_{\rm l,m}=t_{\rm CoFeB,m}=1.3$ nm). The saturation magnetization in thin ferromagnetic layers is usually reduced. For the CoFeB layer of the considered thickness, we assumed the following values of $M_{\rm S}$: $M_{\rm S,m}=1.2\times10^{6}$ A/m  and $M_{\rm S,b}=0.8\times10^{6}$ A/m.~\cite{Devolder16} We also took into account a slight reduction of the exchange stiffness constant in the barrier in reference to its value in the matrix: $A_{\rm m}=27\times10^{12}$ J/m in the matrix region and  $A_{\rm b}=20\times10^{12}$ J/m in the barrier.\cite{Devolder16} To induce the out-of-plane anisotropy, the  CoFeB layer is covered by MgO overlayer. For the strong anisotropy of the CoFeB/MgO interface: $K_{i}=1.3\times10^{-3}$ ${\rm J/m^2}$,\cite{Ikeda10}  both the matrix and barrier are perpendicularly magnetized, with strong effective anisotropy fields: $\mu_0 H_{\rm a,m}=0.16$ T and $\mu_0 H_{\rm a,m}=2.24 T$, respectively. We also assumed an external magnetic field $\mu_0 H_0=0.5$ T applied perpendicularly to the magnetic layer. The corresponding transmissivity and  group delay are shown in Fig.\ref{fig:Fig7} for the barrier width $L=30$ nm. Apart from this, we assumed $t_{\rm mb}$ =2 nm for the width of the interface between the matrix and the barrier. At this interface we assumed $M_{\rm S,mb}$ and $\lambda_{\rm ex,mb}$ corresponding to NBC (see Supplementary Information). In the numerical calculation we used the BMBC with the additional term related to surface anisotropy, which was omitted in analytical considerations and which was irrelevant for the results presented in Figs.\ref{fig:Fig2}-\ref{fig:Fig6}.

In Fig.\ref{fig:Fig7} we present the transmission characteristics for high frequency of SWs passing through the anisotropy barrier formed in the CoFeB layer, as described above. Both the transmissivity (Fig. \ref{fig:Fig7}(a)) and  group delay (\ref{fig:Fig7}(c)) have typical forms and are similar to those presented in Fig.\ref{fig:Fig3}. We were able to adjust the parameters of the model to observe the saturation of the group delay for a relatively narrow barrier, $L=30$ nm. For this width of the barrier we observe noticeable values of $|T|$ in the tunneling regime for $\gamma\mu_0 \omega<H_{\rm a,b}+H_0$ (red part of the plot in Fig.\ref{fig:Fig7}(a)). As a result, the FOM which determines the observation possibility of the HE is slightly enhanced. For instance,  for the frequency $f=65$ GHz we obtain  FOM=0.16 (cf. Fig.\ref{fig:Fig4}(c,f)). The relatively small number of the oscillations (resonances) of $|T|$ for $\gamma\mu_0 \omega>H_{\rm a,b}+H_0$ (green part of the plot in Fig.\ref{fig:Fig7}(a)) results from a relatively narrow width of the barrier assumed here. The larger amplitude of these oscillations originate from the significant contrast of $M_{\rm S}$ in the matrix and barrier.

Figures~\ref{fig:Fig7}(b,d) illustrate the occurrence of the HE in the considered system. One can note the saturation of the group delay $\tau_{gr}$ for larger widths $L$ of the barrier. The saturation is  slower (faster) for higher (lower) frequencies, similarly as in Fig.\ref{fig:Fig5}(c,d). By a careful inspection of  Fig.\ref{fig:Fig7}, one finds that the group delay does not approach zero for $L\rightarrow 0$ (see the red dashed lines). This behavior can be understood when we take into account  the fact that the influence of the interface anisotropy field, present in the BMBC, survives in the limit of $L\to 0$, and gives rise to the nonzero values of  $\tau_{gr}$ in this limit.  To get a uniform system in the limit of $L\to 0$, and thus also a vanishing group delay,
one should simultaneously adjust the interface anisotropy to that in the matrix.

In the numerical studies the saturation value of $\tau_{gr,\infty}$  was approximated by the group delay $\tau_{gr}$ calculated for an extremely wide barrier: L=200 nm (plotted in Fig.\ref{fig:Fig7}(d)). The further extension of the barrier practically does not change $\tau_{gr}$. The dependence of $\tau_{gr,\infty}$ on the frequency, presented in Fig.\ref{fig:Fig7}(d), is qualitatively similar to that in Fig.\ref{fig:Fig6}(b).

\section*{Summary}

In this paper, we have analyzed the Hartman effect for high-frequency (few tens of GHz) spin waves tunneling through a narrow (few tens of nm) magnetic barrier. We investigated a planar system where the barrier was formed by the local increase of a  perpendicular magnetic anisotropy. Such increase may appear due to specific fabrication of the system -- we propose to change the thickness of CoFeB layer covered by MgO to modifiy spatially the effective out-of-plane anisotropy field. The interesting extension of our studies would be the investigations of the system where the anisotropy in the barrier can be tuned on demand. This can be achieved owing to the magnetostriction effect where, with the aid of piezoelectric transducers, the height of anisotropy barrier can adjusted be in a certain range by the application an external electric field \cite{Ota16}.

By calculating the spin wave transmissivity, we have determined the group delay and found its saturation with increasing barrier width. This proves the existence of the Hartman effect for spin waves. We discussed the impact of exchange boundary conditions (determined by the exchange  coupling between the barrier and surrounding material  $A_{\rm mb}$) on the group delay and Hartman effect. We found that decrease of $A_{\rm mb}$ results in the shortening of group delay for tunneling spin waves by the cost of the reduction of the transitivity.

One should  mention that for an exchange dominated system the shape of the transmitted and reflected wave packets is remarkably modified, which follows from the frequency dependence of the transmissivity. This problem can be solved when we consider the tunnelling of dipolar spin waves which can be characterized by linear dispersion relation in range of small wave vectors. However, in dipolar regime we are limited to lower operating frequencies and  restricted to larger sizes of the devices (due to the increase of wavelength).

\bibliography{sample}

\section*{Acknowledgements }
The  study has  received  financial  support  from  the  National  Science
Centre of Poland Grants
No.: UMO-2016/21/B/ST3/00452, UMO-2017/24/T/ST3/00173 and the EU’s Horizon 2020 Research and Innovation Program under Marie Sklodowska--Curie Grant Agreement No. 644348 (MagIC). J.W.K., J.R., Y.S.D. and N.N.D. would like to acknowledge the support of the Foundation of Alfried Krupp Kolleg Greifswald, the Adam Mickiewicz University Foundation and  the Ministry of Education and Science of the Russian Federation: State  Contract No. 3.7614.2017/9.10 and project No. 14.Z50.31.0015, respectively. The authors would like to thank Prof.~M.~Krawczyk for remarks and discussion.

\section*{Author Contributions}
J.B. and I.L.L. brought the idea to consider the Hartman effect for spin waves and proposed the theoretical model for study of this phenomenon, J.W.K did most of theoretical work, Y.S.D, J.R, N.N.D. analyzed the results and preformed the numerical calculations.  All authors participated in discussion and reviewed the manuscript.

\section*{Additional Information}
\textbf{Competing interests:} The authors declare no competing interests.

\section*{S1. Boundary conditions for exchange spin waves}

The Landau-Lifshitz equation (LLE),  describes the dynamics of magnetization $\bm{M}(\bm{r},t)$ in an effective magnetic field. Using this continuous model for the description of spin waves (SWs) propagating  through the interfaces in magnonic systems, we have to define the relevant boundary conditions. The LLE is a second order differential equation (with respect to the spatial coordinates), which requires two boundary conditions in order to determine the integration constants for a general solution. One of the boundary conditions can be found by  integration of LLE in an infinitesimally small surrounding of the interface. The other one has to be postulated using  physical principles, which are not inbuilt in the differential equation itself. One of such principles is the conservation of (exchange) energy flux passing through the interface. For a sharp interface between two magnetic materials and in the absence of any interfacial effects (interfacial anisotropy, arbitrary change of the exchange coupling), the boundary conditions (called \textit{natural boundary conditions} (NBC)) for the amplitudes of the dynamical components of the magnetization $\bm{m}=[m_x,m_y,0]$ (or $m_{+}=m_{x}+im_{y}$ and $m_{-}=m_{x}-im_{y}$) can be formulated in the following form\cite{Klos13,Kruglyak17,Kruglyak14}:
\begin{eqnarray}
\left.\tilde{M}_{\rm{S,l}}^{-1}m_{\delta}(x)\right|_{x=x_0^-}&=&\left.\tilde{M}_{\rm{S,r}}^{-1}m_{\delta}(x)\right|_{x=x_0^+},
\\
\left. \tilde{M}_{\rm{S,l}}\lambda_{\rm{ex ,l}}^2\frac{d m_{\delta}(x)}{dx}\right|_{x=x_0^-}&=&\left. \tilde{M}_{\rm{S,r}} \lambda_{\rm{ex, r}}^2\frac{d m_{\delta}(x)}{dx}\right|_{x=x_0^+},\nonumber\\\label{eq:BC}
\end{eqnarray}
where $x_0=0,L$ are the positions of the interfaces between the matrix and  barrier. The indices $\rm \{l,r\}=\alpha$ denote the material parameters on the left hand side (for $x<x_0$) or right hand side ($x>x_0$) of the interfaces, respectively, which can correspond either to the matrix or to the barrier regions . The index $\delta=\{x,y\}$ or $\{+,-\}$ refers to the different components of dynamical magnetization,

By using the NBC, we assume that the magnetic materials are exchange-coupled at the interface without taking into account interfacial effects. Therefore, the following issues need to be answered: how to include the change of an exchange coupling at the interface?; what is actually the {\it natural} exchange coupling?; when the natural boundary conditions can be applied? These problems were  studied in the 70-ties\cite{Hoffmann70a} (and discussed more extensively in the 90-ties\cite{Pashaev91,Cochran92}) by introducing the interface exchange energy term and the related contribution to effective field, both for the lattice and continuous models. A new type of boundary conditions was introduced by  Hoffmann\cite{Hoffmann70}, which are referred to as \textit{Hoffmann boundary conditions} (HBC). The HBC can be derived on the base of the physical requirement of the continuity of energy flux through the interface\cite{Cochran92,Kruglyak14,Kruglyak17}, which is also fulfilled by the NBC. These boundary conditions, however, fail in the limit of strong exchange coupling at the interface (including the case of a homogeneous medium, for which the materials on both sides of the interface become identical). The corrected boundary conditions introduced by Barna\'s~\cite{Barnas92} and Mills~\cite{Mills92} are referred to as {\it Barna\'s-Mills boundary conditions} (BMBC)\cite{Kruglyak14}. The latter conditions have been used in our paper to calculate the spin wave transmissivity through the barrier of anisotropy field. The general form of HBC and BMBC can also include the impact of different anisotropy fields on both sides of the interfaces. This is especially important in our case where the barrier is formed by anisotropy field.

The BMBC can be written in the following form, which reflects their relation to the NBC and HBC. The first equation reads:
\begin{eqnarray}
\left.D_{{\rm l},\beta}m_{\delta}(x)\right|_{x=x_0^-}&=&\left.D_{{\rm r},\beta}m_{\delta}(x)\right|_{x=x_0^+},\label{eq:BC1}
\end{eqnarray}
where the  operators $D_{l,\beta}=\left(D_{{\rm l},\beta}^{(1)}+D_{{\rm l},\beta}^{(2)}\right)$ and $D_{{\rm r},\beta}=\left(D_{{\rm r},\beta}^{(1)}-D_{{\rm r},\beta}^{(2)}\right)$ are expressed by:
\begin{eqnarray}
D_{\alpha,N}^{(1)}&=&2\frac{\tilde{M}_{\rm{S,mb}}^{2}\lambda_{\rm{ex,mb}}^{2}}{\tilde{M}_{\rm{S,\alpha}}},\nonumber\\
D_{\alpha,N}^{(2)}&=&0,\label{eq:BC_C_N}\\
D_{\alpha,H}^{(1)}&=&D_{\alpha,N}^{(1)}+t_{\rm l,\alpha}\;t_{\rm mb}\;\tilde{H}_{\rm a,\alpha},\nonumber\\
D_{\alpha,H}^{(2)}&=&D_{\alpha,N}^{(2)}+t_{\rm mb}\;\tilde{M}_{\rm{S,\alpha}}\lambda_{\rm{ex,\alpha}}^{2}\frac{d}{dx},\label{eq:BC_C_H}\\
D_{\alpha,BM}^{(1)}&=&D_{\alpha,H}^{(1)},\nonumber\\
D_{\alpha,BM}^{(2)}&=&D_{\alpha,H}^{(2)}-t_{\rm mb}\frac{\tilde{M}_{\rm{S,mb}}^{2}\lambda_{\rm{ex,mb}}^{2}}{\tilde{M}_{\rm{S,\alpha}}}\frac{d}{dx},\label{eq:BC_C_BM}
\end{eqnarray} where the index $\beta=\{N,H,BM\}$ refers to NBC, HBC, BMBC.
Here, the width of the interface between the matrix and the barrier is denoted by $t_{\rm mb}$, and the introduced interfacial parameter,  $\lambda_{\rm ex,mb}$, is the exchange length in the matrix-barrier interface, whereas  $\tilde{M}_{\rm S,mb}$ is the dimensionless saturation magnetization $M_{\rm S,mb}$:
\begin{equation}
\tilde{M}_{\rm S,mb}=\frac{M_{\rm S,mb}}{H_{0}}.
\end{equation}
 These parameters are related to the interface exchange stiffness constant: $A_{\rm mb}=\tfrac{\mu_0}{2}\lambda_{\rm ex,mb}^{2}M_{\rm S,mb}^{2}/t_{\rm mb}$. In turn, the bulk material parameters used here, $\lambda_{\rm ex,\alpha}$ and $\tilde{M}_{\rm S,\alpha}=M_{\rm S,\alpha}/H_{0}$, can be expressed by the bulk exchange  stiffness constant:  $A_{\alpha}=\tfrac{\mu_0}{2}\lambda_{\rm ex,\alpha}^{2}M_{\rm S,\alpha}^{2}$. The parameter $t_{\rm l,\alpha}$ denotes the thickness of the magnetic layer  which is different in the matrix ($\rm \alpha=m$) and in the barrier ($\rm \alpha=b$). The (effective) anisotropy $K_{\alpha}=K_{\rm i}/t_{\rm l,\alpha}-\mu_0 M_{\rm S,\alpha}^2/2$, appearing in the original formulation of the BMBC\cite{Barnas92}, is related here to the efective anisotropy field by the general formula: $\tilde{H}_{\rm a,\alpha}=2K_{\alpha}/(\mu_0 M_{\rm S,\alpha}H_{0})$.

The second equation of the boundary conditions is the same for the NBC, HBC and BMBC:
\begin{eqnarray}
\left.D_{\rm l}m_{\delta}(x)\right|_{x=x_0^-}&=&\left.D_{\rm r}m_{\delta}(x)\right|_{x=x_0^+},\label{eq:BC2}
\end{eqnarray}
where the  operators $D_{\rm l}$ and $D_{\rm r}$  have the form:
\begin{eqnarray}
D_{\alpha}&=&\tilde{M}_{\rm{S,\alpha}}\lambda_{\rm{ex,\alpha}}^{2}\frac{d}{dx}.\label{eq:BC_C}
\end{eqnarray}

By inspection of Eqs.~(\ref{eq:BC1}-\ref{eq:BC_C}), one can notice that for the BMBC,  the components of dynamical magnetization $m_{x}$ and $m_{y}$ are continuous\cite{Barnas92,Kruglyak17}, which is not the case for the HBC. Using the BMBC, we can also correctly determine the values of the interface exchange parameters implicitly existing for the NBC: $A_{\rm mb}/t_{\rm mb}=2A_{\rm m}A_{\rm b}/(A_{\rm m}+A_{\rm b})$ (see Ref.\citeonline{Kruglyak17}) and
\begin{equation}
\lambda_{\rm ex,mb}=\sqrt{2\frac{M_{\rm S,m}M_{\rm S,b}\lambda_{\rm ex,m}^{2}\lambda_{\rm ex,m}^{2}}{M_{\rm S,m}^{2}\lambda_{\rm ex,m}^{2}+M_{\rm S,b}^{2}\lambda_{\rm ex,b}^{2}}},\label{eq:lambda}
\end{equation}
for $M_{\rm S,mb}=\sqrt{M_{\rm S,m}M_{\rm S,b}}$. Moreover, in the range of weak interface exchange coupling (small $A_{\rm mb}$ or $\lambda_{\rm ex,mb}$) the BMBC are reduced to the HBC (see Eq.~\ref{eq:BC_C_BM}).

It is reasonable to assume that thickness of the magnetic layer $t_{\rm l,\lambda}$ (both in barrier and matrix)
is smaller than the width of barrier-matrix interface $t_{\rm mb}$.
 If additionally $t_{\rm l\alpha}$ and $t_{\rm mb}$ are both smaller than  the exchange length ($\lambda_{\rm ex}>t_{\rm mb}>t_{\rm l\alpha}$), then the term  $t_{\rm l,\alpha}t_{\rm mb}\tilde{H}_{\rm a,\alpha}$ in the boundary conditions (\ref{eq:BC_C_N}-\ref{eq:BC_C_BM}) can be neglected. This simplification allows to derive quite clear and compact analytic formulas for the transmissivity and group delay related to the spin wave transmission through the anisotropy barrier.

\section*{S2. The coefficients $\Delta_{\rm s}$ and $\Delta_{\rm c}$ for transmissivity function}
The transmissivity function $T(\Omega,L)$   for SWs tunneling through the anisotropy-field barrier can be written in the form:
\begin{equation}
T(\Omega,L)=\frac{e^{-ik_{\rm m}L}}{\Delta_{c,\beta} \cos(k_{\rm b}L)+i\Delta_{s,\beta} \sin(k_{\rm b}L)}.\label{eq:trans}
\end{equation}
Depending on the boundary conditions, $\beta=\{\rm N,H,BM\}$, the coefficients  $\Delta_{\rm s}$ and $\Delta_{\rm c}$ can be written in the following forms:
\begin{eqnarray}
\Delta_{\rm c,N}&=&1,\nonumber\\
\Delta_{\rm s,N}&=&-\frac{a^2+b^2}{2\,a\,b},\label{eq:deltaN}\\
\Delta_{\rm c,H}&=&\Delta_{\rm c,N}-i\,\frac{a}{c},\nonumber\\
\Delta_{\rm s,H}&=&\Delta_{\rm s,N}+\frac{a\,b}{c^2}+i\,\frac{b}{c}\label{eq:deltaH},\\
\Delta_{\rm c,BM}&=&\Delta_{\rm c,H}+i\,\frac{d}{b},\nonumber\\
\Delta_{\rm s,BM}&=&\Delta_{\rm s,H}+\frac{\,d^{2}}{2\,a\,b}-\frac{d}{c}-i\,\frac{d}{a}.\label{eq:deltaBM}
\end{eqnarray}
The parameters $a=a(\Omega)$ and $b=b(\Omega)$ are the bulk parameters, which depend on $M_{\rm S,\alpha}$ and $\lambda_{\rm ex,\alpha}$;
$c$ is expressed only by the interfacial parameters $M_{\rm S,mb}$,  $\lambda_{\rm ex,mb}$ and $t_{\rm mb}$;  $d=d(\Omega)$  depends on the bulk parameters $M_{\rm S,\alpha}$ and $\lambda_{\rm ex,\alpha}$, and on width of the interface $t_{\rm mb}$:
\begin{eqnarray}
a(\Omega)&=&k_{\rm m}\tilde{M}_{\rm S,m}^{2}\lambda_{\rm ex,m}^{2},\\
b(\Omega)&=&k_{\rm b}\tilde{M}_{\rm S,b}^{2}\lambda_{\rm ex,b}^{2},\\
c&=&\frac{1}{t_{\rm mb}}\tilde{M}_{\rm S,mb}^{2}\lambda_{\rm ex,mb}^{2},\\
d(\Omega)&=&t_{\rm mb}\frac{k_{\rm b}a+k_{\rm m}b}{2}\label{eq:par_d},
\end{eqnarray}
where
\begin{equation}
k_{\alpha}(\Omega)=\lambda_{\rm{ex,\alpha}}^{-1}\tilde{M}_{\rm{S,\alpha}}^{-\frac{1}{2}}\sqrt{\Omega-\left(1+ \tilde{H}_{\rm{a,\alpha}} \right)}.
\end{equation}
The index $\alpha=\{m,b\}$ denotes matrix or barrier.

\end{document}